\documentclass[useAMS,usenatbib]{mn2e} 
\usepackage{graphicx} 
\usepackage{epstopdf} 
\usepackage{amsmath} 
\title[]{Accretion flows with comparable radiation and gas pressures } 
\author[Maryam Samadi, Shahram Abbassi and Wei-Min Gu]{ 
Maryam Samadi $^{1, 2}$\thanks{E-mail:samadimojarad@um.ac.ir}, Shahram Abbassi $^{1}$ \thanks{E-mail:abbassi@um.ac.ir}, Wei-Min Gu $^{3}$\thanks{E-mail:guwm@xmu.edu.cn}\\ 
$^{1}$Department of Physics, School of Sciences, Ferdowsi University of Mashhad, Mashhad, 91775-1436, Iran\\ 
$^{2}$Research Institute for Astronomy and Astrophysics of Maragha (RIAAM)-Maragha, IRAN, P. O. Box:55134 - 441\\ 
$^{3}$Department of Astronomy, Xiamen University, Xiamen, Fujian 361005, China} 
\date{} 
\begin{document} 
\pagerange{\pageref{firstpage}--\pageref{lastpage}} \pubyear{2018} 
\maketitle \label{firstpage} 
\begin{abstract} 
By taking into account photon absorption, we investigate the vertical structure of accretion flows with comparable radiation and gas pressures. We consider two separate energy equations for matter and radiation in the diffusion limit. In order to solve the set of radiation hydrodynamic equations in steady state and axisymmetric configuration, we employ self-similar technique in the radial direction. We need the reflection symmetry about the mid-plane to find gas density at the equator. For a typical solution, we assume that the gas pressure has 10-50\% portion of the total pressure. In this paper, since the radiation energy is involved directly, we are able to estimate how much energy of viscous heating is transported in the
radial direction and advected towards the central object.  Our results show that although the mass accretion rate does not approach the Eddington limit, the energy advection is rather high. Moreover, in a disc with greater accretion rate and less portion of gas pressure at the total pressure, more energy is advected to its center. In addition, as we expect the accretion flow becomes thicker with greater values of gas pressure. Based on Solberg-Høiland conditions, we notice that the flow is convectively stable in all parts of such a disc.  
 
\end{abstract} 
\section{INTRODUCTION} 
In active galactic nuclei (AGN), the major part of energy originates from the vicinity of the central black hole which is occupied by an optically thick accretion flow. Any realistic accretion flow has sufficient angular momentum to form a rotationally supported disc. Due to the liberation of gravitational energy and the existance of turbulent viscosity, the disc can become visible for us. Modeling of such a disc which surrounds a black hole started by pioneers like Zeldovich (1964), Salpeter (1964) and Zeldovich \& Novikov (1967). Lynden-Bell (1969) explained that the 
energy source of quasars is a viscous accretion disc. The standard accretion disc model proposed by Shakura \& Sunyaev (1973) became very successful to describe optically thick and geometrically thin accretion discs observed in X-ray binaries. However, this model turned out unable to clarify some observational feauters arising from very hot plasma around black holes. 
After that popular model, slim disc model were introduced by Abramowicz et al. (1988), presented the properties of those optically thick discs with high mass accretion rates close to its critical values. A few years before it, another popular model had been gradually forming (Ichimaru 1977, Rees et 
al. 1982 ), and finally became comprehensive and identified as radiatevely inefficient accretion flows (RIAFs) (Narayan \& Yi 1994, 1995a,b; Abramowicz et al. 1995; Chen et al. 1997). This model was applied to explain very hard radiation, such as hard X-rays and gamma-rays (up to $100keV$) which were not perceptible by the standard disc model. 
In the slim discs and RIAFs, the heat energy released by viscous dissipation is stored in the gas and it is advected onto the disc's center. Therefore, they both belong to advection-dominated accretion flows and can be appropriately described by the self-similar solutions in the radial direction (Narayan \& Yi 1994). On the other hand, the presence of radiation can influence significantly the dynamics of the accretion system. Moreover, we expect the radiation pressure exceeds the gas pressure in the central regions, since the accretion rate is greater than about 1 $\%$ of the Eddington value (Turner 2004; Shakura \& Sunyaev 1973; Pringle 1981). 
It seems that without including radiation pressure in the definition of $\alpha-$prescription, we cannot explain observed X-ray luminosities of some compact objects. Near Eddington limit, photons carry a significant fraction of the total momentum, hence radiation also associates with matter for transportation of the angular momentum out of the disc (Jiang et al. 2013). 
So far, some simulations have been done by several athours (e.g. Eggum et al. 1987, 1988; Okuda et al. 1997; Fujita \& Okuda 1998; Kley \& Lin 1999; Takeuchi et al. 2009; Sadowski et al. 2014) to study slim discs especially in supercritical accretion regime. Fujita \& Okuda (1998) performed 2D calculations to examine subcritical accretion discs with a luminosity comparable with Eddington one. Okuda (2002) noticed that it is possible for a flow which has supercritical accretion rate in the outer boundary layers to become subcritical near the black hole. An analytical investigation for a time-independent case has been done by Gu (2012) (Gu12) which was drawn attention to the vertical structure of such a flow with dominant radiation pressure. In the pioneer theoretical works of Shakura-Sunyaev and slim disks, it is common to consider asymptotic states of the accreting systems which one of the two pressures of gas or radiation is dominant and also they assume just electron scattering has the main role in the opacity of disc. In the standard model, in the region close to the central mass, the radiation pressure reaches to its ultimate limit, and the high amount of luminosity comes out from this part of disc. Further that luminous part, the main assumption changes to the other ultimate limit of pressure where the gas pressure becomes more significant than radiation one. Therefore, we do not expect these ultimate limits occur between two completely discrete neighborhood regions, it does not sound unreasonable to have an expectation to find a transition area between these two territories with both considerable pressures of photons and real particles. On the other hand, we think of another probable situation which is related to being an independent area extended from $r=10 r_g$ to several tens of Schwarzschild 
radius with $p_r\approx p_g$. Another point related to luminosity of the mentioned region might come to our mind: whether the high luminosity can also be produced by a part of the flow where both gas and radiation have comparable portions and a non-negligible number of photons are trapped and therefore this might affect the flow's dynamics. Moreover, one of basic assumptions of the standard disc is the local energy balance between viscous heating and cooling via radiation but a bit deviation from it leads to extra energy for advection. If we consider conservation energy of matter and radiation separately, we will be able to investigate the variation of energy advection at the intermediate area with mass accretion rate. 
 
 In this paper, we follow the similar procedure of Gu12 and employ self-similar solutions to remove $r$- dependency of the system quantities but we aim to study the mentioned different situation where both radiation and gas have comparable portions to form the total pressure of system. 
The accretion disc's thickness has been examined by Gu et al. (2009) with assumption of semi-polytropic relationship, $p(\theta)=K\rho(\theta)^\gamma$, where $p, \rho$ and $\gamma$ are gas pressure, density and the ratio of specific heats. The important result of that work was to show that the accretion flows in advection-dominated regime tend to be quite thick. In our earlier studies (Samadi et al. 2014 and Samadi \& Abbassi 2017), we focused on the effect of magnetic field with toroidal and poloidal configurations, separately. For both cases, the magnetic force acted in the opposite direction of the gas pressure force and made the disc thinner. In the current work, we might expect the radiation pressure behaves like gas pressure and decreases towards the polar axis. 
The main purpose of this paper is to revisit vertical structure of optically thick accretion flows with rather high accretion rates (saying, several tens of percent of Eddington mass accretion rate) by taking into account photon absorption. The outline of this paper is as follows. In Section 2 we present the basic radiation hydrodynamics equations, which include the two separate equations for energies of gas and radiation. 
Self-similar equations are employed for a typical radius of $r=30 r_g$ in Section 3. Section 4 is devoted to boundary conditions and the permissible values of density at the equatorial plane regarding the gas temperature. The results of the numerical solution and a derivation of the disc thickness are presented 
in Section 5 and, finally, discussions and conclusions are given in Section 6. 
\section{Equations and assumptions} 
We study an accreting flow around a black hole with mass $M$, in steady-state regime which is assumed to be axisymmetric and non-relativistic. We define the system in spherical coordinates,$(r,\theta,\phi)$ and we use Newtonian potential, that is, $\Psi=GM/r$. 
The basic equations to describe such a system, are consist of 
\newline 
1. continuity equation, 
\begin{equation} 
\frac{\partial\rho}{\partial t}+\nabla\cdot(\rho\textbf{v})=0, 
\end{equation} 
where $\rho$ is density and $\textbf{v}$ is the time-averaged flow velocity. \newline 
2. the momentum conservation, 
\begin{equation} 
\rho\frac{D\textbf{v}}{Dt}=-\nabla p_g-\rho\nabla\Psi-\nabla\cdot\textbf{T}^\nu+\frac{\chi}{c}\textbf{F}_0 
\end{equation} 
where $D/Dt=\partial/\partial t+\textbf{v}.\nabla$, and, $p_g$ is gas pressure, $\textbf{T}^\nu$ is stress tensor, $c$ is light speed,$\textbf{F}_0$ is the radiation flux. The total opacity $\chi$ has two parts including Thomson scattering and absorption 
\begin{equation} 
\chi=\frac{\sigma_T}{m_p}\rho+\kappa 
\end{equation} 
where $\sigma_T=6.652\times 10^{-25}\hspace*{0.2cm}cm^2$ is the Thomson cross-section, $m_p$ is the proton mass and $\kappa$ is the absorption opacity. In this problem, we consider two kinds of absorption, saying free-free and bound-free, as 
\begin{displaymath} 
\kappa_{ff}=1.7\times 10^{-25} T^{-7/2}(\frac{\rho}{m_p})^2\hspace*{0.2cm}cm^{-1}, 
\end{displaymath} 
\begin{displaymath} 
\kappa_{bf}=4.8\times 10^{-24} T^{-7/2}\frac{Z}{Z_\odot}(\frac{\rho}{m_p})^2\hspace*{0.2cm}cm^{-1}, 
\end{displaymath} 
So we can find the total absorption opacity as $\kappa=\kappa_{ff}+\kappa_{bf}$. 
\newline 
3. the energy equation of gas, 
\begin{equation} 
\frac{\partial e}{\partial t}+\nabla\cdot(e\textbf{v})=-p_g\nabla\cdot\textbf{v}+\Phi_{vis}-4\pi\kappa B+c\kappa E_0, 
\end{equation} 
where $e$ is the internal energy density, $\Phi_{vis}$ is the viscous dissipative function, $B$ is the blackbody intensity and $E_0$ is the radiation energy density. \newline 
4. and finally based on conservation of radiation energy density (Mihalas \& Mihalas 
1984) we have, 
\begin{equation} 
\frac{\partial E_0}{\partial t}+\nabla\cdot(E_0\textbf{v})=-\nabla\textbf{v:}\textbf{P}_0-\nabla\cdot\textbf{F}_0+4\pi\kappa B-c\kappa E_0, 
\end{equation} 
where $\textbf{P}_0$ is the radiation pressure tensor. 
It might be helpful to point out the physical interpretation of new terms which have appeared in Eq.(5) due to considering radiation and also the interaction between photons and gas particles. For instance, $\nabla\textbf{v:}\textbf{P}_0$ shows the rate at which radiation pressure is doing work, $\nabla\cdot\textbf{F}_0$ specifies the rate of transport of radiation energy, $4\pi\kappa B$ is the rate at which energy of matter is being added to the radiation field (the same term in Eq.(4) appears with negative sign that means decreasing energy of matter due to its radiation) and finally $-c\kappa E_0$ denotes the amount of radiation energy which is removed from the total radiation energy and transported to the energy of matter in the unit of time (all terms are per unit volume). 
In the spherical coordinates, the velocity field has these three components, $( v_r, v_\theta, v_\phi)$ but we follow Narayan \& Yi (1995), and assume $v_\theta=0$\footnote{This assumption is often used for all kinds of accretion discs.}. With steady-state, axisymmetric flow and neglecting $v_\theta$, Eq. (1) yields, 
\begin{equation} 
\frac{1}{r^2}\frac{\partial}{\partial r}(r^2\rho v_r)=0, 
\end{equation} 
\newline 
For simplicity, we use Eddington approximation for the radiation tensor, that is, 
\begin{equation} 
\text{\textbf{P}}_{0ij} = 
\begin{cases} 
p_r=E_0/3 & \text{if}\hspace*{0.3cm} i=j \\ 
0 & \text{if}\hspace*{0.3cm} i\neq j 
\end{cases} 
\end{equation} 
this means that $\textbf{P}_0$ behaves like a scalar and simply equals to radiation pressure, $p_r$. 
The radiation flux vector, $\textbf{F}_0= (F_{0r}, F_{0\theta}, F_{0\phi})$, in diffusion approximation has a relationship with gradient $E_0$ as $\textbf{F}_0=-c\nabla E_0/(3\chi)$, using Eq.(7) we will find: 
\begin{displaymath} 
\textbf{F}_0=-\frac{c}{\chi}\nabla p_r, 
\end{displaymath} 
Therefore, the components of radiation flux become, 
\begin{equation} 
F_{0r}=-\frac{c}{\chi}\frac{\partial p_r}{\partial r},\hspace*{0.3cm} F_{0\theta}=-\frac{c}{\chi}\frac{\partial p_r}{r\partial \theta},\hspace*{0.3cm} F_{0\phi}=0, 
\end{equation} 
Now we can find three scalar equations based on Eq.(2) as, 
\begin{equation} 
v_r\frac{\partial v_r}{\partial r}-\frac{v_\phi^2}{r}=-\frac{GM}{r^2}-\frac{1}{\rho}\frac{\partial}{\partial r}(p_g+p_r), 
\end{equation} 
\begin{equation} 
v_\phi^2\cot\theta=\frac{1}{\rho}\frac{\partial}{\partial\theta}(p_g+p_r) 
\end{equation} 
\begin{equation} 
v_r\frac{\partial}{\partial r}(r v_\phi)=\frac{1}{\rho r^2}\frac{\partial}{\partial r}(r^3 T_{r\phi}) 
\end{equation} 
where we have assumed that the $r\phi-$ component of the viscous stress tensor is dominant and we neglect the other components of $T^\nu$, 
\begin{equation} 
T_{r\phi}=\eta r\frac{\partial}{\partial r}\big(\frac{v_\phi}{r}\big) 
\end{equation} 
where $\eta$ is the dynamical viscosity coefficient. We apply $\alpha$-prescription (Shakura \& Sunyaev 1973) to determine the dynamical viscosity, means 
\begin{equation} 
\eta=\alpha\frac{p_g+p_r}{\Omega_K} 
\end{equation} 
where $\Omega_K$ is the Keplerian angular velocity. 
To determine the internal energy density, we can use, 
\begin{equation} 
e = \frac{p_g}{\gamma -1} 
\end{equation} 
The last term in Eq. (4) that we need to know is 
the heating rate of viscosity, $\Phi_{vis}$, given by 
\begin{equation} 
\Phi_{vis}=\eta \big[r\frac{\partial}{\partial r}\big(\frac{v_\phi}{r}\big)\big]^2. 
\end{equation} 
Moreover, $B$ is the blackbody intensity and defined as, 
\begin{equation} 
B = \frac{\sigma T^4}{\pi}, 
\end{equation} 
where $T$ is the gas temperature and $\sigma$ is Stefan-Boltzmann constant. To calculate the gas temperature, we can use the following relationship, 
\begin{equation} 
T=\frac{\mu m_p}{k_B}\frac{p_g}{\rho}, 
\end{equation} 
where $k_B$ is the Boltzmann constant and is the mean molecular 
weight; in the present study we assume $\mu=0.5$. 
Now substituting Eq.(13)-(17) in Eq.(4) besides our basic assumptions, we obtain, 
\begin{displaymath} 
\frac{1}{r^2}\frac{\partial}{\partial r}(\frac{ p_g}{\gamma-1} r^2v_r)=-\frac{p_g}{r^2}\frac{\partial}{\partial r}(r^2v_r)-4\kappa\sigma T^4, 
\end{displaymath} 
\begin{equation} 
\hspace*{0.3cm}+3c\kappa p_r+\alpha\frac{p_g+p_r}{\Omega_K}\big[r\frac{\partial}{\partial r}\big(\frac{v_\phi}{r}\big)\big]^2, 
\end{equation} 
Using Eq.(7), (8), (15) and (16) in the last basic equation of (5) leads us to find, 
\begin{displaymath} 
\frac{1}{r^2}\frac{\partial}{\partial r} (3r^2 p_r v_r)=\frac{c}{r^2}\frac{\partial}{\partial r}\bigg( \frac{r^2}{\chi}\frac{\partial p_r}{\partial r}\bigg)+\frac{c}{r\sin\theta}\frac{\partial}{\partial\theta}\bigg(\frac{\sin\theta }{\chi}\frac{\partial p_r}{r\partial \theta}\bigg) 
\end{displaymath} 
\begin{equation} 
-\bigg(\frac{\partial v_r}{\partial r}+2\frac{v_r}{r}\bigg)p_r+4\kappa\sigma T^4-3c\kappa p_r, 
\end{equation} 
(refer to appendix to see the details of calculation of $\nabla\textbf{v}\textbf{:}\textbf{P}_0$). 
\begin{figure*} 
\centering 
\includegraphics[width=160mm]{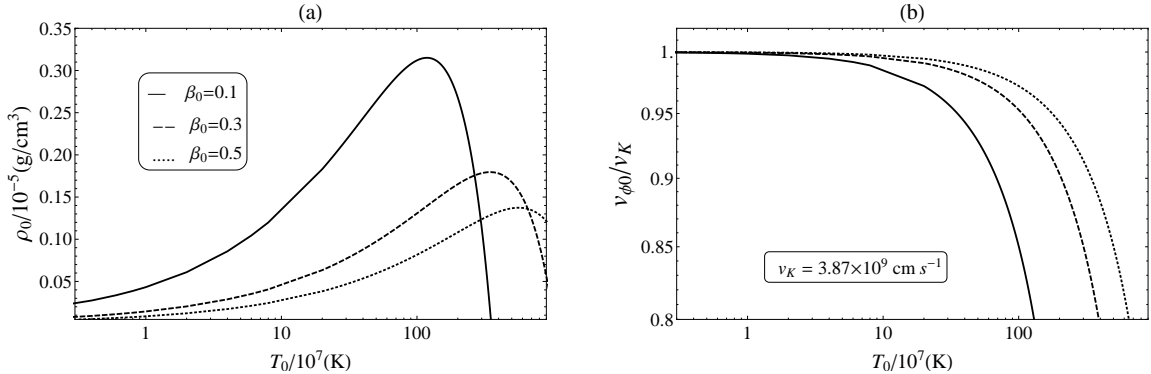} 
\caption{Solutions at the equatorial plane corresponding to $M=10M_\odot, \alpha=0.3, \gamma=1.5$ and several values of $\beta_0$. The radius of these solutions is $r=30r_g$, where $r_g=2GM/c^2$ is the Schwarzschild radius. Here, $v_K$ is Keplerian velocity and equals to $3.87\times 10^9 cm s^{-1}$.} 
\end{figure*} 
\begin{figure*} 
\centering 
\includegraphics[width=185mm]{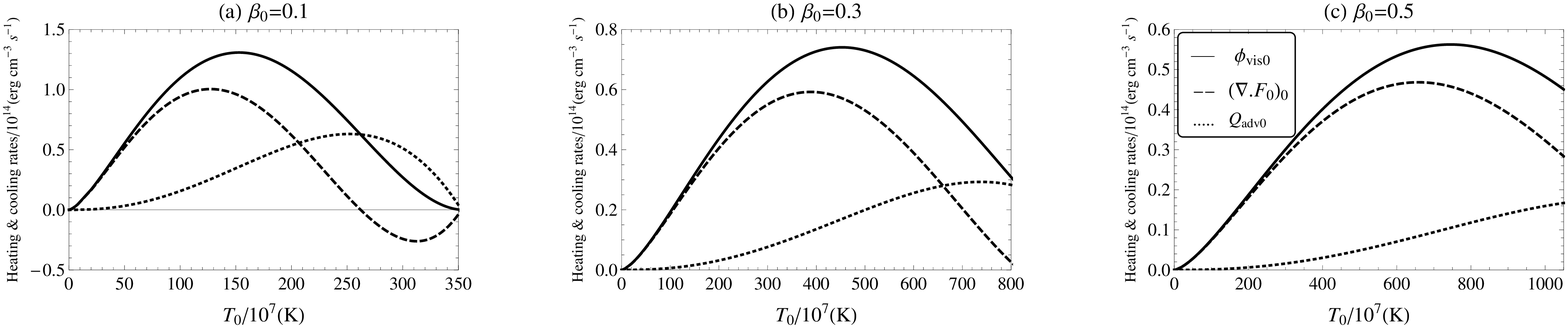} 
\caption{The heating ($\Phi_{vis0}$) and cooling ($Q_{adv0}$, $(\nabla\cdot\textbf{F}_0)_0$) rates at the equatorial plane corresponding to $r=30r_g$, $M=10M_\odot, \alpha=0.3, \gamma=1.5$ and several values of $\beta_0$. } 
\end{figure*} 
\section{Self-similar solutions} 
In this work, we are interested in studying vertical structure of the disc with comparable pressures of gas and radiation, so we need to use self-similar technique in the radial direction to remove $r$- dependency of the quantities and find equations with respect to just one variable, means $\theta$. Based on self-similarity we know the dependency of quantities are power law as, 
\begin{equation} 
v_i\propto r^{-1/2},\hspace*{0.3cm}\rho\propto r^{-3/2}, \hspace*{0.3cm} p_i\propto r^{-5/2}, \hspace*{0.3cm} T\propto r^{-1},\hspace{0.3cm} \textbf{F}_{0i}\propto r^{-2} 
\end{equation} 
Employing these solutions in Eq. (9)-(11), (18) and (19) gives, 
\begin{equation} 
v_\phi^2=v_K^2-\frac{v_r^2}{2}-\frac{5}{2\rho}(p_g+p_r), 
\end{equation} 
\begin{equation} 
\rho v_\phi^2\cot\theta=\frac{d}{d\theta}(p_g+p_r), 
\end{equation} 
\begin{equation} 
v_r=-\frac{3\alpha}{2\rho v_K}(p_g+p_r), 
\end{equation} 
\begin{equation} 
\frac{1}{2} v_r (3\rho v_\phi^2-\frac{5-3\gamma}{\gamma-1}p_g)=3rc\kappa[p_r-\frac{4\sigma}{3c}T^4 ] 
\end{equation} 
\begin{equation} 
\frac{3}{2}v_r p_r+\frac{c}{r}(\cot\theta+\frac{d}{d\theta})[\frac{1}{\chi}\frac{dp_r}{d\theta}]=3rc\kappa[p_r-\frac{4\sigma}{3c}T^4] 
\end{equation} 
Now, we have six equations (17) and (21-25) including six unknowns ($T, \rho, v_r, v_\phi, p_r$ and $p_g$) that we should find their dependencies to $\theta$. In order to solve the system of ODEs, we should first find boundary conditions that comes in the following. 
\section{Solutions at the equator} 
So far, we have obtained a set of six equations that only two of them are differential equations, means Eq.(22) and (25). From these three algebraic Eq. (17), (21) and (23-24) which consist of $T, \rho, v_r, v_\phi, p_g$ and $p_r$, so we can determine only four unknowns. Since we are somewhat familiar with the temperature of gas in optically thick discs, we treat this quantity as a free parameter. We also apply $\beta_0$ parameter (zero index denotes the value of each quantity at $\theta=\pi/2$) as the ratio of gas pressure to the total pressure at the equatorial plane, 
\begin{displaymath} 
\beta_0=\frac{p_{g0}}{p_{g0}+p_{r0}}, 
\end{displaymath} 
It is clear that for radiation pressure supported discs we expect $\beta_0\sim 0$ and for gas pressure dominated discs we have $\beta_0\sim 1$. 
Here, it is convenient to fix the equatorial value of gas temperature and calculate the corresponding $p_r$ and then find the proper density for several $\beta_0$. The solutions of the three-term algebraic equations system are displayed in figures 1 and 2 for $M=10M_\odot$, $\mu=0.5$ at the radius $r=30r_g$, where $r_g(=2GM/c^2)$ is the Schwarzschild radius. According to Fig.(1), the maximum permissible density at the equator roughly reaches $3\times 10^{-6}\ g\ cm^{-3}$ for $T_0\approx 10^9\ K$ and $\beta_0=0.1$. Moreover, the maximum available density at the mid-plane becomes smaller with greater $\beta_0$. 
The variation of rotational velocity at $\theta=90^\circ$ in the second panel of Fig.1 is presented and shows that 
$v_{\phi0}$ in fairly cold discs (with temperature less than $10^8K$) is approximately Keplerian and in discs with greater radiation pressure it drops more rapidly. 
Here, we need to take care of the sign of $\nabla\cdot\textbf{F}_0$, because it is expected to be positive and show the transportation of energy outwards. The first panel of Fig.(2) reveals that the mentioned term becomes zero at $T_0=3.5\times 10^9 K$ for $\beta_0=0.1$. Comparing the second and third panels with the first one, we notice that with greater $\beta_0$, the radiation cooling transforms to heating term at larger gas temperatures. At the limit of low temperatures of all panels, the viscous heating is equal to the radiation cooling ($\Phi_{vis0}=(\nabla\cdot\textbf{F}_0)_0$) and energy advection is zero. 
\begin{figure*} 
\centering 
\includegraphics[width=180mm]{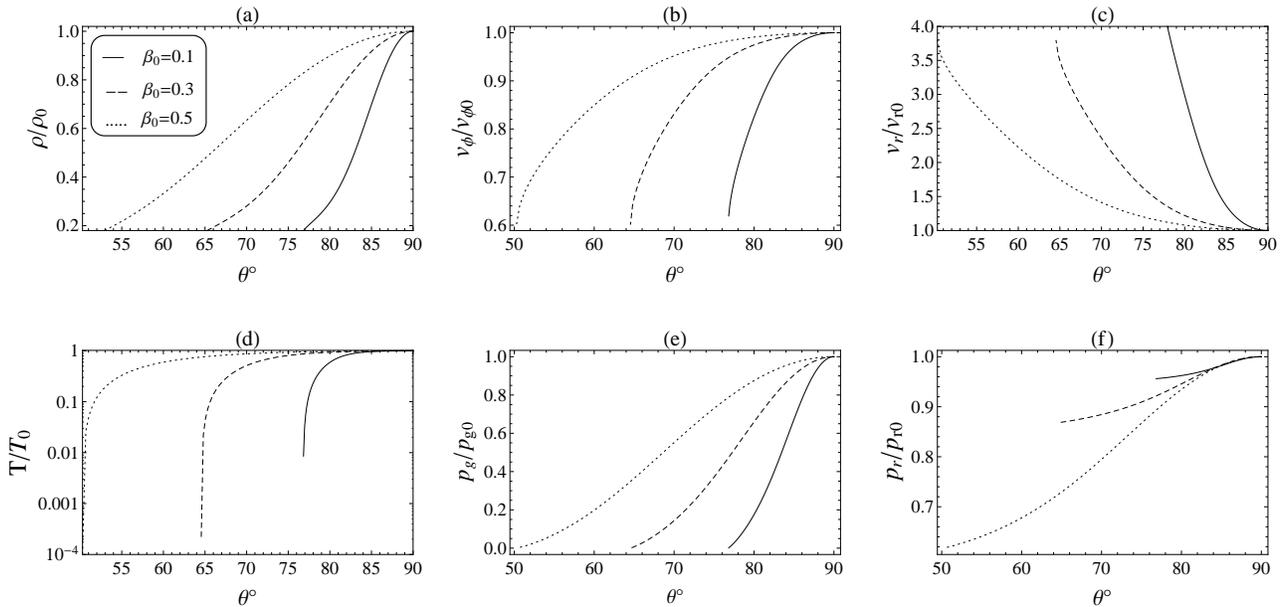} 
\caption{Angular profile of the physical variables 
at $r=30r_g$ corresponding to $\alpha=0.3, \gamma=1.5, M=10M_\odot, \beta_0=0.1$ and several values of $\beta_0$.} 
\end{figure*} 
\begin{figure} 
\centering 
\includegraphics[width=70mm]{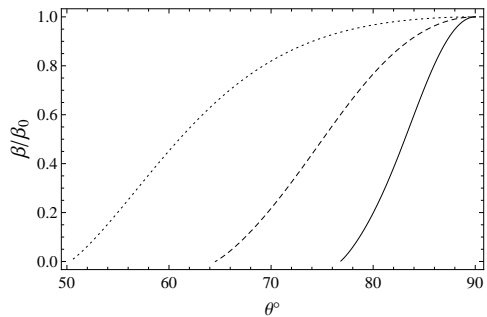} 
\caption{The angular variation of $\beta$'s at $r=30r_g$, which shows how much the ratio of gas pressure to the total pressure of disc reduces towards the surface and tends to even zero at the outer layers. The solid, dashed 
and dotted lines refer to $\beta_0 = 0.1, 0.3$ and $0.5$, respectively. } 
\end{figure} 
\section{Vertical Structure} 
In this section, we refer to the two differential equations of our system, i.e.. equations (22) and (25). For solving Eq.(22), we need to specify one boundary condition. From the previous section, density and gas and radiation pressures corresponding to temperature and a certain value of $\beta_0$ are known at the equatorial plane. Similarly for Eq.(25) which is second order ODE, two boundary conditions are required. Here beside $\rho_0, T_0, p_{g0}$ and $p_{r0}$, we employ the reflection symmetry about the disc's equator to assume the derivatives of density and pressures are equal to zero, means 
\begin{displaymath} 
\theta=90^\circ:\hspace*{0.5cm}\rightarrow \frac{d\rho}{d\theta}=\frac{dp_g}{d\theta}=\frac{dp_r}{d\theta}=0 
\end{displaymath} 
We start the integrations of ODEs from the equatorial plane and stop our calculations when gas pressure becomes zero. For these sets of input parameters: $M=10M_\odot, \alpha=0.3, \gamma=1.5$ and for $\beta_0=0.1, 0.3, 0.5$ with these temperatures: $T_0/10^8 K=0.5, 2, 5$ at $r=30r_g$, the integrations might be continued and end at the poles for some other cases. With smaller gas temperature like $5\times 10^8 K$ that we have used for $\beta_0=0.1$, we stop integrations at $\theta_s\approx 77^\circ$ (we call $\theta_s$, the surface angle) because the gas pressure, $p_g$ becomes zero at this angle. 
\begin{figure*} 
\centering 
\includegraphics[width=160mm]{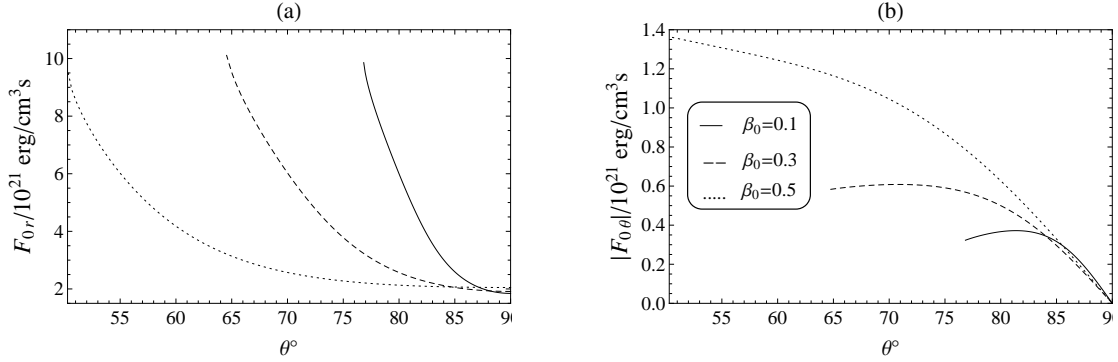} 
\caption{Angular profile of radiation flux's components for a typical solution at $r=30r_g$ with $\alpha=0.3, \gamma=1.5, M=10M_\odot$ and three different values of $\beta_0$.} 
\end{figure*} 
\begin{figure*} 
\centering 
\includegraphics[width=180mm]{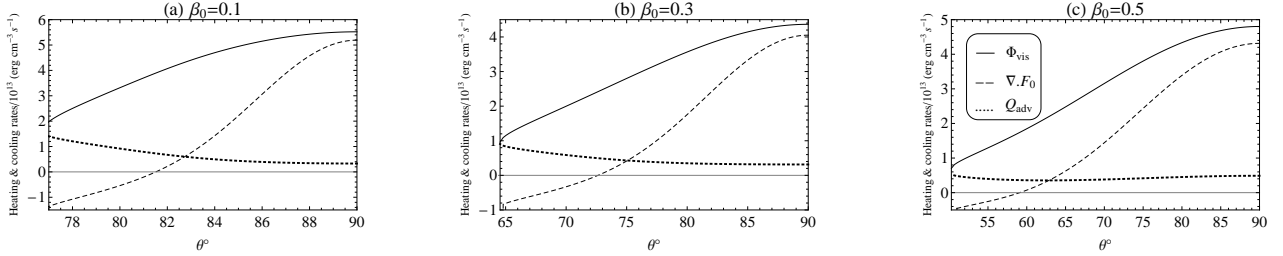} 
\caption{The angular variation of heating ($\Phi_{vis}$) and cooling ($Q_{adv}$, $\nabla\cdot\textbf{F}_0$) rates at $r=30r_g$ with $\alpha=0.3, \gamma=1.5$ and $M=10M_\odot$ with different $\beta_0$' s.} 
\end{figure*} 
Figure (3) shows the angular variation of dimensionless density, radial and azimuthal components of the flow's velocity and also the two pressures of gas and radiation. 
All quantities in this figure have been scaled by their boundary values at the equatorial plane which are listed in table (1). 
According to Fig.(3a), the density of gas starts from a maximum value at the mid-plane and tends to its minimum at the surface angle. Comparing the minimum value of our density here with most of those in ADAFs we have faced with in our previous works, we notice this minimum cannot reach to zero and just becomes about 20\% of its equatorial value. However, this result is not strange for theoretical solutions, since similar little ascending trend can be seen in Fig.3 Jiao \& Wu 2011 (the minimum value of $\rho$ is about $0.4\rho_0$ and also in Fig.1 of Narayan \& Yi 1995a, $\rho$ varies by only 10\% from the polar axis to the mid-plane with $\epsilon'=0.1$ and for $\epsilon'=1$ we see $\rho(\theta=90^\circ)\sim 2\rho(\theta=0)$. In profile (b), it can be seen that the rotational velocity is decreasing outwards, whereas the accretion velocity is minimum at the mid-plane and ends up maximum at the surface. The gas temperature, $T$ (equivalently the sound speed squared $c_s^2=k_B T/\mu m_p$ ) peaks at $\theta=90^\circ$, that is the same as ADAF's solutions in other works (such as Fig.1,4 of Samadi et al. 2013, Fig.1 of Gu et al. 2009). The two pressures appear somehow different, $p_r$ does not show noticeable change inside the disc especially for lower value of $\beta_0$, i.e.. 0.1. As we expect for gas pressure, it faces with a rather sharp decrease towards the surface of disc. 
\begin{table} 
\caption{Equatorial values of quantities} 
\centering 
\begin{tabular}{c rrr} 
\hline 
\hline 
$\beta_0$  & 0.1  & 0.3  & 0.5 \\ 
\hline 
\hline 
$T_0/10^9 (K)$  & 0.5 & 2 & 5 \\ 
\hline 
$\rho_0/10^{-6} (g.cm^{-3})$ & 2.64 & 1.65 & 1.37 \\ 
\hline 
$v_{\phi0}/v_K$  & 0.93 & 0.90 & 0.85 \\ 
\hline 
$-v_{r0}/v_K$  & 0.02 & 0.03 & 0.05\\ 
\hline 
$p_{g0}/10^{11}\ (dyn. cm^{-2})$ & 2.18 & 5.46 & 11.29\\ 
\hline 
$p_{r0}/10^{12}\ (dyn. cm^{-2})$  & 1.96 & 1.27 & 1.13\\ 
[1ex] 
\hline 
\end{tabular} 
\label{tab:hresult} 
\end{table} 
\begin{table} 
\caption{Results} 
\centering 
\begin{tabular}{c rrr} 
\hline 
\hline 
$\beta_0$  & 0.1 & 0.3  & 0.5 \\ 
\hline 
\hline 
$\dot{m}$ & 0.20 & 0.28 & 0.44 \\ 
\hline 
$\Delta\theta^\circ$  & 13.17 & 25.46 & 39.72 \\ 
\hline 
H/r & 0.23 & 0.43 & 0.64 \\ 
\hline 
$f_{adv}$ & 0.15 & 0.14 & 0.13 \\ 
\hline 
$Q_{vis}/10^{21}\ (erg\ cm^{-3} s^{-1})$ & 0.86 & 1.17 & 1.80 \\ 
\hline 
$Q_{adv}/10^{20}\ (erg\ cm^{-3} s^{-1})$  & 1.33 & 1.72 & 2.37 \\ 
\hline 
$Q_c/10^{20}\ (erg\ cm^{-3} s^{-1})$  & 3.14 & 5.25 & 10.5\\ 
[1ex] 
\hline 
\end{tabular} 
\label{tab:hresult} 
\end{table} 
To estimate boundary conditions we have set $\beta_0=0.1$ and according to Fig.4, $ 
\beta$ is not constant and goes down and will tend to be zero at the surface, thus radiation is dominant at the outer layers. 
Here we can substitute the density and radial velocity in the following equation and evaluate the mass accretion rate, 
\begin{equation} 
\dot{M}=4\pi r^2\int_{90^\circ}^{\theta_0}\rho v_r\sin\theta d\theta 
\end{equation} 
at $r=30r_g$ with $\beta_0=0.3$, we find $\dot{M}=0.20\dot{M}_{Edd}$ (see table 2 for two other values of $\beta_0$) that $\dot{M}_{Edd}$ is Eddington accretion rate and has this definition: $\dot{M}_{Edd}=4\pi GM/(\eta c\kappa_{esc}$, where $\eta$ is the 
radiative efficiency of the flow and we chose $\eta = 1/16$ since it is 
comparable to the Schwarzschild black hole efficiency of 0.057 (Gu12). 
In the presence of radiation flux, the flow's dynamics is affected by the two effective components of radiation force. According to Fig.(5), the both components, radial $F_r=+5 cE_0/(6\chi r)$ and meridional $F_\theta=-c /(3\chi r)dE_0/d\theta $ are directed out of the flow. Moreover, the absolute value of the radiation flux force is increasing outwards in the current case. 
An important aspect of hot accretion flows is their ability to store some of dissipated energy (via turbulent viscosity) as entropy and it is advected with the flow. This advection energy can be written in the following form, 
\begin{equation} 
q_{adv}=\rho T\frac{dS_t}{dt}=\rho\frac{d}{dt}(\frac{e+E_0}{\rho})-\frac{p_t}{\rho}\frac{d\rho}{dt}, 
\end{equation} 
where $S_t$ and $p_t$ are the total entropy and total pressure, respectively. Using the definition of Lagrangian time derivative, Eq.(15) and also assumptions of steady-state, $v_\theta=0$, self-similar radial solutions in Eq.(27), we obtain 
\begin{equation} 
q_{adv}= 
-\frac{5-3\gamma}{2(\gamma-1)}\frac{p_gv_r}{r}-\frac{3p_rv_r}{2r}, 
\end{equation} 
The advection energy besides two other energies of $\Phi_{vis}$ and $\nabla\cdot\textbf{F}_0$ are plotted in Fig.6 for three values of $\beta_0$. For all these $\beta_0$'s, we see that the amount of $\Phi_{vis}$ is larger than $q_{adv}$ and divergence of radiation flux. As mentioned before, in common solutions the sign of $\nabla\cdot\textbf{F}_0$ is expected to be positive and show the cooling term. This expectation is satisfied in the typical solution presented here but not all over the regions of the accretion flow as seen in Fig.6 and it becomes zero at $81.5^\circ$ for the disc with$\beta_0=0.1$. We speculate some reasons for this issue. First of all, the radiation energy should be transported outside, along the vertical direction over the photosphere, here with just $\theta-$dependency 
of quantities, we can be sure about agreement between our results in spherical coordinates and those in cylindrical one from the equatorial plane up to just the small distances from the mid-plane. On the other hand, we have ignored $v_\theta$ and this causes $v_r$ remains negative and even more negative towards the rotation axes. The negative value of $v_r$ besides self-similarity solutions leads us to evaluate the left-hand side of both Eq.(4) and (5) with positive sign ($ev_r\propto r^{-3}\rightarrow \nabla\cdot(e\textbf{v})=1/r^2 d/dr(r^2 e v_r)=-ev_r/r>0$) which means that both the internal gas and radiation density energy are increasing with time in the fixed frame. The main difference between these two equations is that the viscous heating assure us about the presence of a source to increase $e$ but in Eq.(5) instead of that, there is $-\nabla\cdot\textbf{F}_0$ which is expected to be negative and act as a cooling term. At regions close enough to the equatorial plane, the last two terms in Eq.(5) by exchanging energy between particles and photons compensate for lack of immediate or direct source of heating to enhance the radiation energy density. The other reason for negative sign of $\nabla\cdot\textbf{F}_0$ might be unsatisfactory of self-similar technique near the surface of the disc. Despite being negative of divergence $\textbf{F}_0$, the meridional component of radiation flux remains in the same direction as before according to the right panel of figure 5. 
Equation (28) says that the advection energy depends on both pressures of gas and radiation and radial velocity. Although both gas and radiation pressures are minimum at the surface, the profile of advection energy $q_{adv}$ in Fig.6 does not diminish because of considerable growth in the absolute value of radial velocity. The total energy of advection ($Q_{adv}$), viscosity ($Q_{vis}$) and cooling ($Q_c$) at a certain radius can be calculated as, 
\begin{displaymath} 
Q_{adv}=2\int^{\theta_0}_{90^\circ}q_{adv} r\sin\theta d\theta, 
\end{displaymath} 
\begin{displaymath} 
Q_{vis}=2\int^{\theta_0}_{90^\circ}\Phi_{vis} r\sin\theta d\theta, 
\end{displaymath} 
\begin{displaymath} 
Q_{c}=2\int^{\theta_0}_{90^\circ}\nabla\cdot\textbf{F}_0 r\sin\theta d\theta, 
\end{displaymath} 
For the typical solution at $r=30r_g$ with $\beta_0=0.3$, we find $Q_{adv}=1.33\times 10^{20}\ erg\ cm^{-3} s^{-1}$, $Q_{vis}=8.65\times 10^{20}\ erg\ cm^{-2} s^{-1}$ and $Q_c=3.14\times 10^{20}\ erg\ cm^{-3} s^{-1}$ (see table 2 for two other values of $\beta_0$). Therefore, the vertically averaged advection factor becomes, 
\begin{equation} 
f_{adv}=\frac{Q_{adv}}{Q_{vis}}, 
\end{equation} 
which is approximately equal to $0.15$. 
\begin{figure*} 
\centering 
\includegraphics[width=180mm]{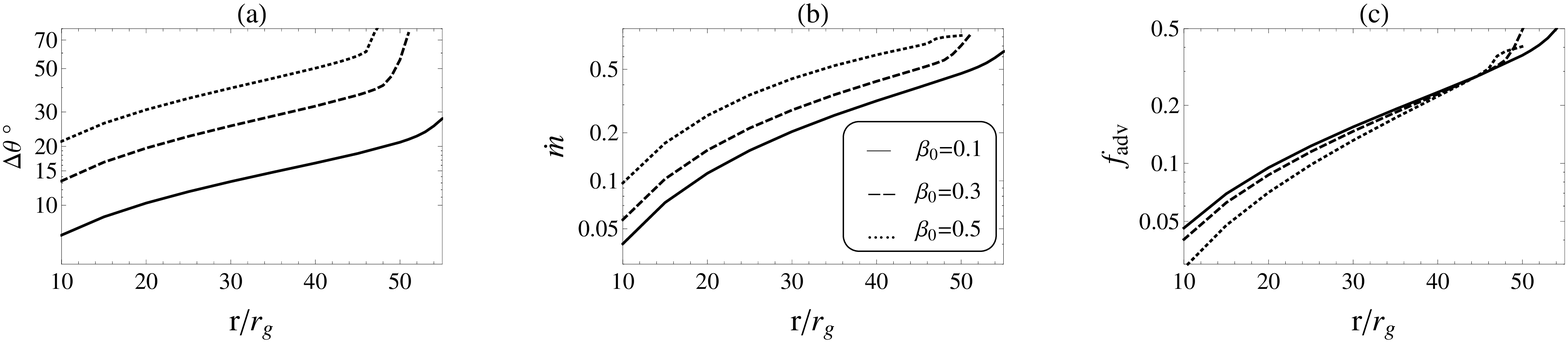} 
\caption{Variation of (a) the disc's half-thickness, (b) mass accretion rate and (c) advection parameter with respect to radius. The equatorial value of gas temperature and density are different for each $\beta_0$. } 
\end{figure*} 
In figure (7), we have plotted (a) the half-thickness of disc (b) the total mass accretion rate and (c) advection parameter versus radius for three values of $\beta_0$. As seen in panels of this figure, all these quantities increase at larger radii. Furthermore, in the last figure of this section, fig.8, we see a direct effect from mass accretion rate on the advection factor. The similar result have been presented by Gu (2012) in figure 4 of his paper. 
\begin{figure} 
\centering 
\includegraphics[width=70mm]{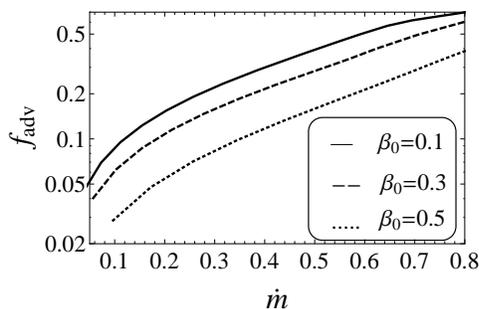} 
\caption{Variation of the advection parameter, $f_{adv}$ with respect to the mass accretion rate, $\dot{m}$.} 
\end{figure} 
\subsection{Convective Stability} 
The last thing which would be our interest to examine is the convection instability in the optically thick accretion discs with comparable gas and radiation pressures. To do this, we should refer to Solberg-Høiland conditions for dynamical stability which are defined in cylindrical coordinates $(R,\phi,z)$ as following, 
\begin{equation} 
\frac{1}{R^3}\frac{\partial j^2}{\partial R}-\frac{1}{C_p\rho}\nabla p_t\cdot\nabla S>0 
\end{equation} 
\begin{equation} 
\frac{\partial p_t}{\partial z}(\frac{\partial j^2}{\partial R}\frac{\partial S}{\partial z}-\frac{\partial j^2}{\partial z}\frac{\partial S}{\partial R})<0 
\end{equation} 
where $j=R v_\phi=r\sin\theta v_\phi$ is the specific angular momentum per unit mass and the differential of entropy is defined by $dS=d\ln(p_t/\rho^\gamma)$. On the other hand, the oscillation frequencies related to the internal hydrodynamic modes are driven by equilibrium gradients. In this way, the square of the epicyclic frequency is obtained as, 
\begin{equation} 
\kappa^2=\frac{1}{R^3}\frac{\partial j^2}{\partial R} 
\end{equation} 
\begin{figure*} 
\centering 
\includegraphics[width=185mm]{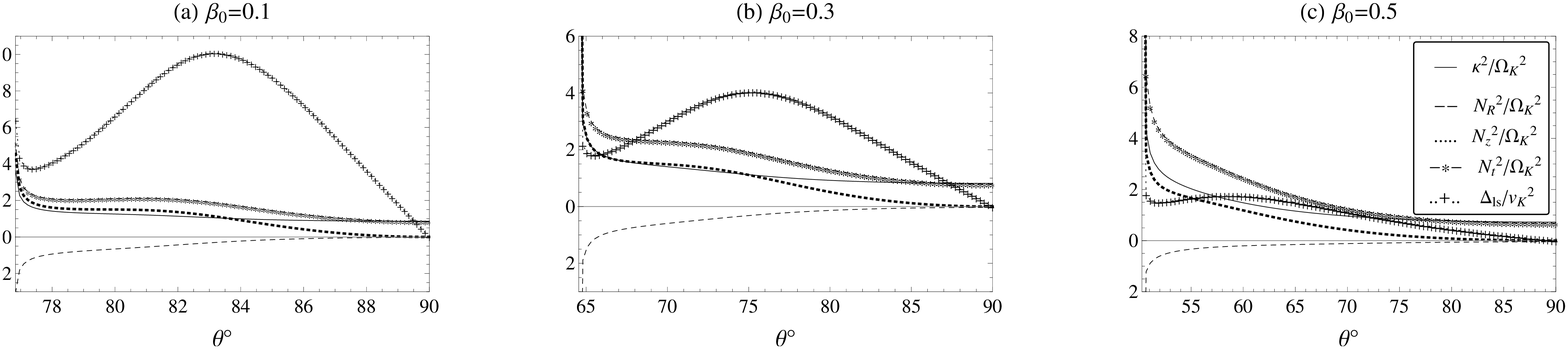} 
\caption{The angular variation of $\kappa^2$ (thick solid line), $N_R^2$ (dashed line), $N_Z^2$ (dotted line), $N_t^2$ (dashed line with asterix) and $\Delta_{ls}$ (dotted line with plus sign) at $r=30r_g$ with $M=10M_\odot, \alpha=0.3, \gamma=1.5$ and for three values of $\beta_0$.} 
\end{figure*} 
The two components of the Brunt–Väisälä frequency are given by 
\begin{displaymath} 
N_R^2=-\frac{1}{\gamma\rho}\frac{\partial p_t}{\partial R}\frac{\partial}{\partial R}\ln(\frac{p_t}{\rho^\gamma }) 
\end{displaymath} 
\begin{displaymath} 
N_z^2=-\frac{1}{\gamma\rho}\frac{\partial p_t}{\partial z}\frac{\partial}{\partial z}\ln(\frac{p_t}{\rho^\gamma }) 
\end{displaymath} 
Substituting the last three equations in Eq.(31) gives the first condition of a convectively stable flow as, 
\begin{equation} 
N_t^2=\kappa^2+N_R^2+N_z^2>0 
\end{equation} 
where we have named the total frequency as $N_t^2$. Furthermore, the second condition is simplified as 
\begin{equation} 
\Delta_{ls}=\frac{\partial j^2}{\partial R}\frac{\partial}{\partial z}\ln(\frac{p_t}{\rho^\gamma})-\frac{\partial j^2}{\partial z}\frac{\partial}{\partial R}\ln(\frac{p_t}{\rho^\gamma})>0 
\end{equation} 
as seen, we have omitted the factor $\partial p_t/\partial z$ because this term is usually negative in accretion discs, in addition our solutions of $p_r$ and $p_g$ in figure 3 
are in agreement with it. Since we have found the angular dependency of the quantities, it is convenient to transform derivatives with respect to $R$ and $z$ into the spherical coordinates, by means of the chain rule, 
\begin{displaymath} 
\frac{\partial}{\partial R}=\sin\theta\frac{\partial}{\partial r}+\frac{\cos\theta}{r}\frac{\partial}{\partial\theta} 
\end{displaymath} 
\begin{displaymath} 
\frac{\partial}{\partial z}=\cos\theta\frac{\partial}{\partial r}-\frac{\sin\theta}{r}\frac{\partial}{\partial\theta} 
\end{displaymath} 
By using the self-similar solutions and our results from the previous section we have calculated these frequencies and shown in figure 9. The radial component of Brunt–Väisälä frequency, $N_R^2$ is negative and the other component of it is positive. As it is seen in fig.9, the total frequency of $N_t^2$ and $\Delta_{ls}$ for all three are positive, thus the current disc is convectively stable. 
\section{Summary and conclusion} 
We studied the vertical structure of optically thick accretion flows in the presence of both radiation and gas pressures. We did not employ radiative transfer, instead of that we adopted the separated energy equation similar to the internal energy of gas based on diffusion approximation in radiation field. The self-similar solutions in the radial direction helped us to convert the partial differential equations into a set of algebraic equations and ODE's. 
Solving equations on the equator and considering some basic assumptions like high optical depth, we could obtain proper values of density and gas temperature. At the typical radius of $r=30r_g$, we assumed the gas pressure had 10-50\% portion of the total pressure and calculated the $\theta-$dependency of physical quantities. 
With a permissible gas temperature at mid-plane ($T_0/10^9 K=0.5, 2, 5$), we calculated corresponding gas pressure which yields $\beta_0=0.1, 0.3, 0.5$. The velocity's components at the equatorial plane were found from two algebraic equations. With these known quantities, we could start integrating the differential equations and then stopped calculation because of gas pressure which tended to negative values. In fact, the gas pressure showed a descending behaviour and it became gradually zero at a certain angle ($0<\theta_0<90$) before approaching the vertical axis, whereas the radiation pressure faced with a little change at this angular distance. Therefore, we expect to find accretion discs with two vertical parts, one is located inside from the mid-plane to $\theta_0$, and affected by both $p_g$ and $p_r$, and the other one as a surrounding layer of the equatorial part whose radiation pressure is dominant and hence it has different features. Although the temperature of gas at the inside region was rather high at the equator, it became negligible at the surface of the disc, meanwhile the temperature at the outer layer is determined based on radiation and it must be very high. Furthermore, we found out with greater accretion rate and less portion of gas pressure in the total pressure, more energy was advected to the central object.
In our solutions, we realized that despite of the outward direction of radiation flux, it is possible that divergence of the radiation flux to become negative. However, our results for other input parameters 
are not practical at outer regions (especially for hotter discs) because we neglected the meridional component of the velocity and it is just ignorable near the mid-plane but $v_\theta$ is very effective to make $v_r$ positive and change significantly the angular profile of energy transportation outward the flow. Moreover, the typical accretion flows presented here with relatively we saw them convectively stable. 
Our results here are useful to apply for intermediate cases between slim and thin discs which include two parts, one part near the mid-plane which has nonignorable gas pressure and the other part with dominant radiation pressure and negligible gas pressure (about 3 order of magnitude smaller). The vertical structure of inner part can be found by the presented method but it is incomplete. The outer part can be described by omitting the gas pressure and use one energy equation for the total energy of the flow like which has been done by Gu12. We can use the surface value of quantities as boundary conditions for solving differential equations in the outer part. 
\section*{Acknowledgements}
We are grateful to the anonymous referee for his/her thoughtful and
constructive comments, which helped us to improve the first version of this paper.
Maryam Samadi also wishes to thank Faride Danesh-Manesh for her kind helps in some parts of this project. This work has been supported financially by Research Institute for Astronomy \& Astrophysics of Maragha (RIAAM) under research project No. 1/6025-24.

\appendix 
\section{Formulas of Tensor production } 
To simplify the tensor production of $\nabla\textbf{v:}\textbf{P}_0$ in the radiation energy equation (Eq.5) in the spherical coordinate, we present the basic formulas here to calculate it conveniently. The double dot (or scalar, or inner) product produces a scalar, (Phan-Thien 2013) 
\begin{equation} 
\textbf{U}:\textbf{V}=(U_{ij}\textbf{e}_i \textbf{e}_j):(V_{kl}\textbf{e}_k \textbf{e}_l)=U_{ij}V_{ji} 
\end{equation} 
The gradient of a vector is given by 
\begin{displaymath} 
\nabla\textbf{u}=\textbf{e}_r\textbf{e}_r\frac{\partial u_r}{\partial r}+\textbf{e}_r\textbf{e}_\theta\frac{\partial u_\theta}{\partial r}+\textbf{e}_r\textbf{e}_\phi\frac{\partial u_\phi}{\partial r}+\textbf{e}_\theta \textbf{e}_r\bigg(\frac{1}{r}\frac{\partial u_r}{\partial\theta}-\frac{u_\theta}{r}\bigg) 
\end{displaymath} 
\begin{displaymath} 
+\textbf{e}_\theta \textbf{e}_\theta\bigg(\frac{1}{r}\frac{\partial u_\theta}{\partial\theta}+\frac{u_r}{r}\bigg)+\textbf{e}_\theta \textbf{e}_\phi\frac{1}{r}\frac{\partial u_\phi}{\partial\theta}+\textbf{e}_\phi \textbf{e}_r\bigg(\frac{1}{r\sin\theta}\frac{\partial u_r}{\partial\phi}-\frac{u_\phi}{r}\bigg) 
\end{displaymath} 
\begin{equation} 
+\textbf{e}_\phi \textbf{e}_\theta\bigg(\frac{1}{r\sin\theta}\frac{\partial u_\theta}{\partial\phi}-\frac{u_\phi}{r}\cot\theta\bigg)+ 
\textbf{e}_\phi \textbf{e}_\phi\bigg(\frac{1}{r\sin\theta}\frac{\partial u_\phi}{\partial\phi}+\frac{u_r}{r}+\frac{u_\theta}{r}\cot\theta\bigg) 
\end{equation} 
So the production of $\nabla\textbf{v}$ and $\textbf{P}_0$ becomes, 
\begin{displaymath} 
\nabla\textbf{v}\textbf{:}\textbf{P}_0=\frac{\partial v_r}{\partial r}\textbf{P}_{0rr}+\bigg(\frac{1}{r}\frac{\partial v_\theta}{\partial\theta}+\frac{v_r}{r}\bigg)\textbf{P}_{0\theta\theta} 
\end{displaymath} 
\begin{equation} 
+ 
\bigg(\frac{1}{r\sin\theta}\frac{\partial v_\phi}{\partial\phi}+\frac{v_r}{r}+\frac{v_\theta}{r}\cot\theta\bigg)\textbf{P}_{0\phi\phi} 
\end{equation} 
with the assumptions of $v_\theta=0$, $\partial/\partial\phi=0$ and $\textbf{P}_{0ii}=E_0/3$ we have, 
\begin{displaymath} 
\nabla\textbf{v}\textbf{:}\textbf{P}_0=\bigg(\frac{\partial v_r}{\partial r}+2\frac{v_r}{r}\bigg)\frac{E_0}{3} 
\end{displaymath} 
\end{document}